\pdfoutput=1
\documentclass[aps,prl,preprint,groupedaddress,nofootinbib,11pt]{revtex4-1}
\usepackage[]{natbib}
\usepackage{amsmath, amssymb, amsthm, url}
\usepackage{graphicx}
\usepackage[table]{xcolor}  
\usepackage[normalem]{ulem} 
\usepackage{dcolumn}
\usepackage{setspace}
\usepackage{booktabs}
\usepackage{lineno} 
\newcolumntype{d}[1]{D{.}{.}{#1}}

\begin{document}
\title{Phage-antibiotic synergy inhibited by temperate and chronic virus competition}

\author{Kylie J.~Landa}
\affiliation{Department of Mathematics, Statistics, and Computer Science, St.~Olaf College, Northfield, Minnesota 55057, USA}

\author{Lauren M.~Mossman}
\affiliation{Department of Mathematics, Statistics, and Computer Science, St.~Olaf College, Northfield, Minnesota 55057, USA}

\author{Rachel J.~Whitaker}
\affiliation{Department of Microbiology, University of Illinois at Urbana-Champaign, Urbana, Illinois 61801, USA}
\affiliation{Carl R.~Woese Institute for Genomic Biology, University of Illinois at Urbana-Champaign, Urbana, Illinois 61801, USA}

\author{Zoi Rapti}
\affiliation{Department of Mathematics, University of Illinois at Urbana-Champaign, Urbana, Illinois 61801, USA}
\affiliation{Carl R.~Woese Institute for Genomic Biology, University of Illinois at Urbana-Champaign, Urbana, Illinois 61801, USA}

\author{Sara M.~Clifton}
\email{clifto2@stolaf.edu}
\affiliation{Department of Mathematics, Statistics, and Computer Science, St.~Olaf College, Northfield, Minnesota 55057, USA}

\keywords{bacteria, bacteriophage, phage, latent, lytic, infection, recovery, resistance, Pseudomonas aeruginosa, cystic fibrosis, population dynamics, mathematical model}

\date{\today}

\spacing{1.1}

\begin{abstract} 
As antibiotic resistance grows more frequent for common bacterial infections, alternative treatment strategies such as phage therapy have become more widely studied in the medical field. While many studies have explored the efficacy of antibiotics, phage therapy, or synergistic combinations of phages and antibiotics, the impact of virus competition on the efficacy of antibiotic treatment has not yet been considered. Here, we model the synergy between antibiotics and two viral types, temperate and chronic, in controlling bacterial infections. We demonstrate that while combinations of antibiotic and temperate viruses exhibit synergy, competition between temperate and chronic viruses inhibits bacterial control with antibiotics. In fact, our model reveals that antibiotic treatment may counterintuitively \emph{increase} the bacterial load when a large fraction of the bacteria develop antibiotic-resistance. 

\end{abstract}

\maketitle


\section{Introduction}

The rising number of antibiotic resistant bacteria is causing an increase in the cost of treatment, higher mortality rates, and a longer period of sickness \cite{duse2013}. For instance, infections by the multi-drug resistant pathogen \textit{Pseudomonas aeruginosa}, which is categorized as a serious threat by the CDC \cite{CDC2019}, are frequently diagnosed among those with compromised immune systems to such an extent that it is the most common cause of death among patients with cystic fibrosis \cite{hancock2000antibiotic}. 

As antibiotic resistant bacteria are becoming more prevalent, new strategies are needed to combat these bacterial infections. Recent experiments show that antibiotics and phages may synergistically target bacterial populations when used in tandem \cite{chaudhry2017synergy,easwaran2020}. While some studies have revealed that two phage types can synergistically fight bacterial infections \cite{chaudhry2017synergy}, we demonstrate that certain virus types that are naturally present in human hosts may inhibit bacterial infection control with antibiotics. Because the viruses compete for access to host cells and once the cells are infected the viruses exclude one another, thus restricting the sensitive population for the other virus.

The two viral lifestyles studied in this paper are temperate and chronic. When chronic viruses infect a bacterium, the host cell produces and releases new viruses without lysing itself \cite{weinbauer2004ecology}. Chronic viruses also have a latent (lysogenic) cycle in which viral genetic material is incorporated into the bacterium’s genome. This genetic code is then transmitted to daughter cells when the bacterium replicates \cite{lwoff1953lysogeny} . Unlike chronic viruses, temperate viruses have a lytic and a latent cycle. During the latent cycle, the virus remains dormant in the bacterium until induced to replicate. During the lytic cycle, the bacterium produces viruses that burst out of the cell, causing the host to die \cite{weinbauer2004ecology}. This phenomenon of lysis can be triggered by an SOS response induced in the bacterium cell \cite{weinbauer2004ecology}.

The SOS response is an inducible repair system found in bacteria such as \textit{P. aeruginosa}. Stress from a bacterium’s environment such as UV light, sublethal doses of antibiotics, and radiation can precipitate this evolved response \cite{fothergill2011effect,rokney2008host}. When certain classes of antibiotics, such as quinolones, are introduced into the bacteria’s environment, they have the potential to damage the bacterial DNA which can then induce the SOS response \cite{power1992}. Once the SOS response has been activated, bacteria may start to replicate previously latent viruses, potentially leading to cell death \cite{Brives2020}.   

This paper uses a mathematical model to investigate the effects of the two competing virus types, temperate and chronic, on the effectiveness of various antibiotic dosing frequencies. In most cases, we find that as the frequency of antibiotic dosages increases, the bacterial population decreases even if the bacteria are known to be antibiotic resistant. However, our model reveals that antibiotic treatment counterintuitively \emph{increases} the bacterial load when a large fraction of the bacteria develop antibiotic-resistance and temperate and chronic phages are present in the system.

\section{Model}
\begin{figure}[h]
\centering
\includegraphics[width=\textwidth]{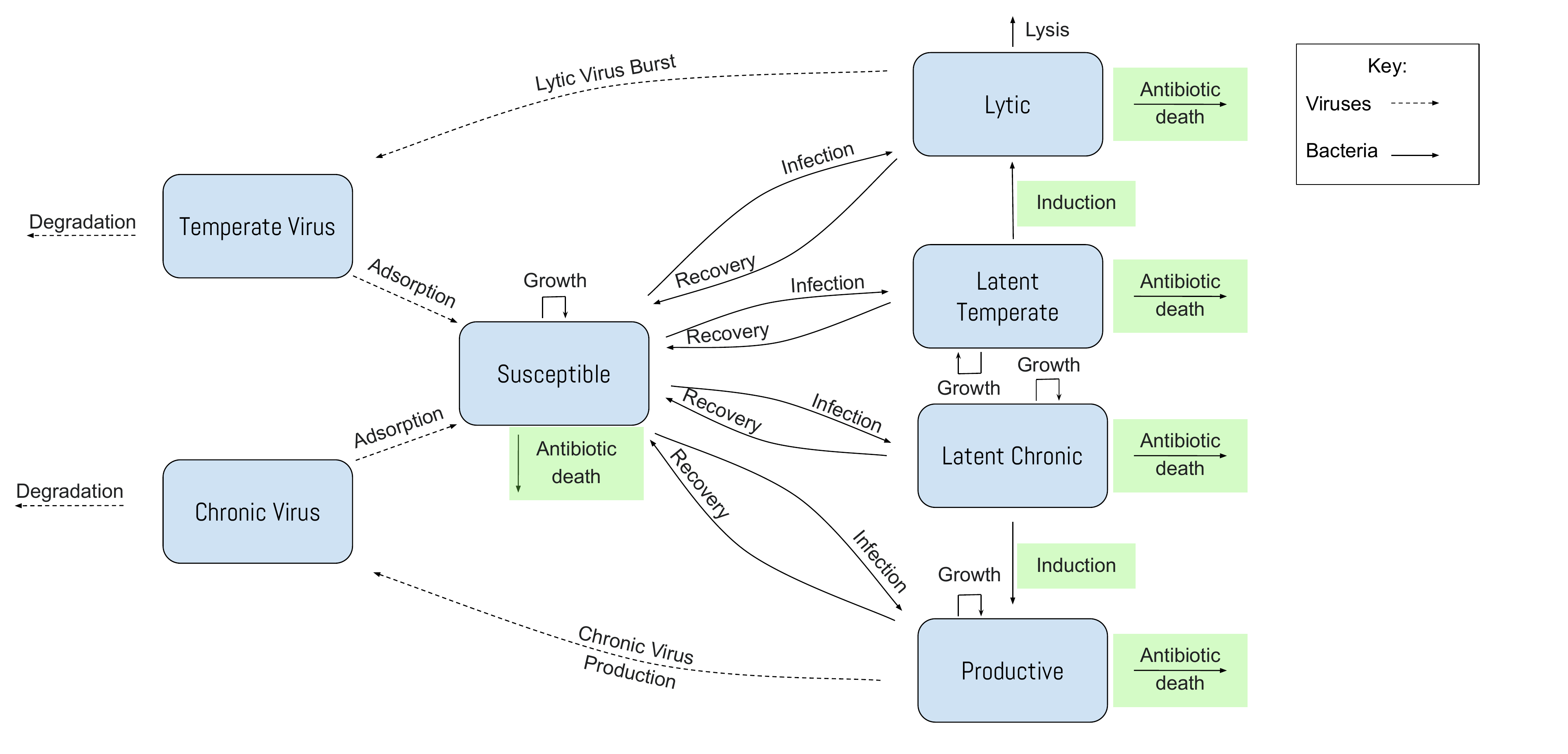}
\caption{Compartmental model of bacteria-phage system with temperate and chronic phages. Solid arrows represent bacteria transitions between infection states, and dotted arrow represent virus production or loss. The green boxes illustrate actions which may occur only when environmental stress, such as sublethal antibiotics, are introduced to the system.} \label{fig:flowchart}
\end{figure}

We begin with a model of temperate and chronic virus competition \cite{clifton2019temperate}, which assumes that two virus types, temperate viruses ($V_T$) with lytic and latent lytic stages and chronic viruses ($V_C$) with productive and latent chronic stages, may infect a single strain of bacteria that is initially susceptible ($S$) to both viral types. See Figure \ref{fig:flowchart} for a model overview.


The total bacterial population is composed of susceptible, lytic, latent temperate, productive, and latent chronic bacteria. We assume that the total bacterial population grows logistically to a carrying capacity $K$ \cite{zwietering1990modeling}. In the model system, we rescale all population densities to be proportional to the carrying capacity, and we rescale the timescale such that all rates are relative to the susceptible bacteria's growth rate.

The susceptible bacteria grow at an intrinsic rate $r_S$ and can become infected by chronic or temperate viruses. The susceptible bacteria are infected by the temperate viruses at a rate $\eta_T$. When infected, there is a $f_T$ probability that the bacterium enters a latent lytic state in which viruses are not being produced but the viral DNA is integrated into the bacterium’s genome and passed on to later generations \cite{harper2014bacteriophages}. The latent lytic bacteria reproduce at a rate $r_T$ and recover from infection at a rate $\gamma_T$. 

Alternatively, there is a $(1 - f_T)$ probability that the bacteria will instead enter the lytic lifestyle following infection. The bacteria hosting phages in a lytic lifestyle die due to lysis (bursting of cell) at a rate $\delta$. When the bacteria lyse, they produces $\beta_T$ temperate viruses per burst bacterium. It is important to note that in both cases of recovery, the bacteria become susceptible to phage infection once again. The bacteria can recover from infection at a rate $\gamma_I$.

Similar to temperate infections, the susceptible bacteria are infected by the chronic viruses at a rate $\eta_C$. When infected, there is a $f_C$ probability that the bacteria will enter the latent chronic state in which the bacteria can reproduce at a rate $r_C$ and recover from infection at a rate $\gamma_C$. 

Alternatively, there is a $(1-f_C)$ probability that the bacteria will instead become productive upon infection and grow at a rate $r_P$ while producing viruses at a rate $\beta_C$. In contrast to lytic infections, productive infections do not cause cell death. Productive bacteria recover from infection at a rate $\gamma_P$. 

Outside the bacterial cells, free temperate viruses and free chronic viruses are adsorbed by susceptible bacteria at a rate $\eta_T$ and $\eta_C$, respectively, and degrade naturally at a rate $\mu_T$ and $\mu_C$, respectively \cite{heldal1991production}. Free temperate viruses are produced when the lytic bacteria burst at a rate $\beta_T \delta$, and free chronic viruses are generated by productive bacteria at a rate $\beta_C$.

Once a bacterium is infected, we assume it will exclude both superinfection by the same viral type and cross infection by viruses of the other type \cite{de2017pseudomonas}. Although we assume infection rates for temperate and chronic viruses are constant, the infection rates may depend on the bacterial population; many relevant bacteria form biofilms at high density that protect the population from infection \cite{harper2014bacteriophages}. For simplicity, we have also assumed that lysogen frequencies are constant, but some studies have demonstrated that bacterial density may impact lysogeny rates \cite{hargreaves2014does,silpe2018host}. These simplifications are necessary for analytic tractability.

\subsection{Antibiotics}
We introduce antibiotics into the system at times $t_i$; specifically these are the times at which the antibiotics become bioavailable, which depends on the mode of delivery. We assume that system stress spikes at times $t_i$ and decays exponentially, consistent with typical antibiotic metabolism in the human system \cite{naber1973pharmacokinetics,bax1989pharmacokinetics, fish1997clinical}. The antibiotic concentration within the system is then:
\begin{align}
	a(t) = A \, \sum_{i=1}^N H(t-t_i) \, e^{-k (t-t_i)},
\end{align}
where $t$ is the current time, $t_i$ are the antibiotic dose times, $A$ is the maximum concentration of one antibiotic dose, $N$ is the total number of antibiotic doses, $H$ is the Heaviside function, and $k$ is the decay rate of antibiotics in the system. 

When the system is stressed by antibiotics, there are three major responses. First, antibiotics can induce the production of latent temperate or chronic viruses at a rate $\epsilon \, a(t)$ \cite{fothergill2011effect, ptashne1986genetic, nanda2015impact}. This response occurs with bacteria that are antibiotic resistant or antibiotic susceptible \cite{fisher2017persistent,monack2004persistent,stewart2001antibiotic,james2012differential, redgrave2014fluoroquinolone,valencia2017ciprofloxacin,brazas2005ciprofloxacin}. However, not all classes of antibiotics induce virus production, so this model is only applicable to antibiotics (such as quinolones) known to do so \cite{zhang2000quinolone,comeau2007phage}. To simplify the model, we ignore spontaneous induction because it is a rare phenomenon \cite{nanda2015impact,garro1974relationship,Cortes546275}. 

Second, antibiotic resistant bacteria are not killed directly by the antibiotics \cite{fisher2017persistent,monack2004persistent,stewart2001antibiotic}, but antibiotic susceptible bacteria are killed at a rate proportional to the antibiotic concentration: $\kappa \, a(t)$ \cite{levin2010population}. 

\begin{figure}[htbp]
  \centering
    \includegraphics[width=0.45\textwidth]{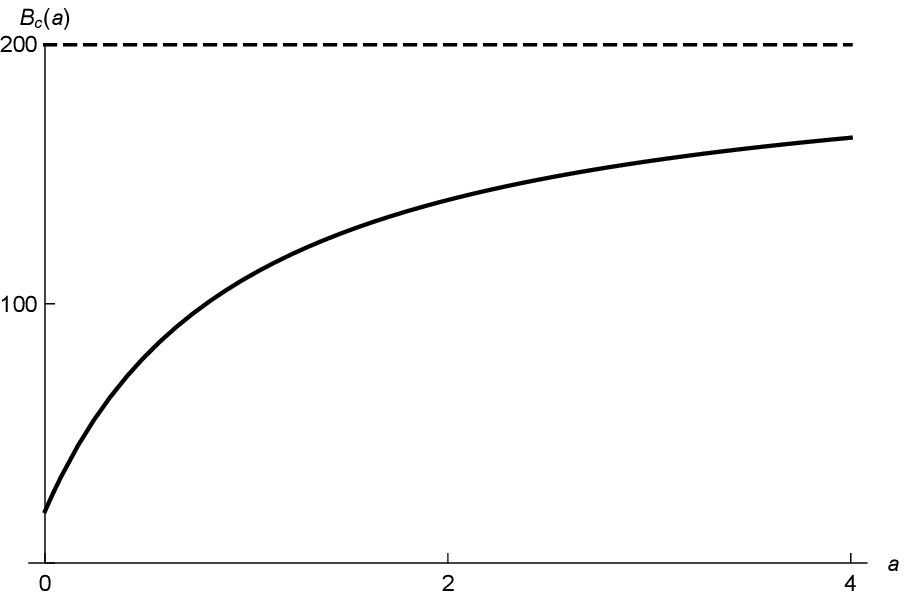}
    \includegraphics[width=0.45\textwidth]{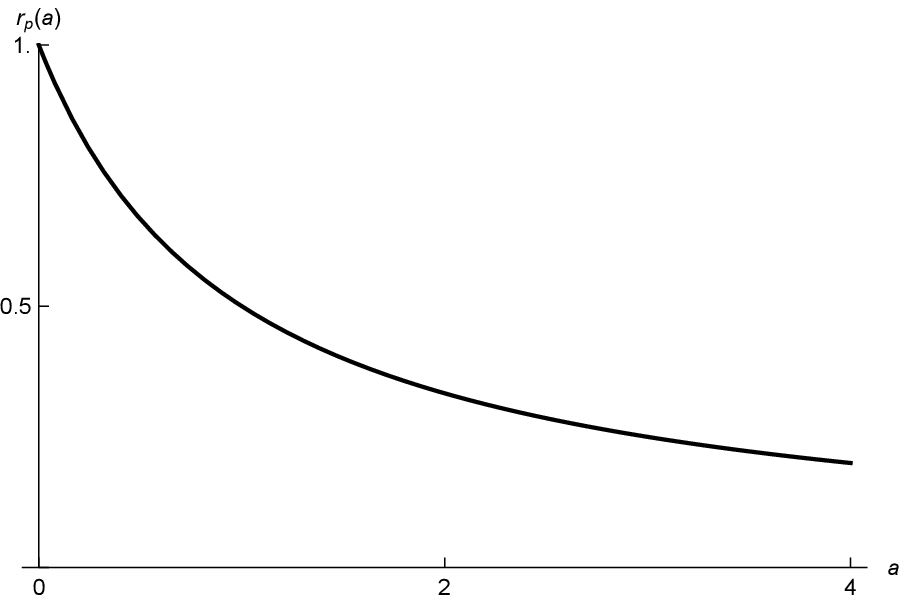}
      \caption{Functional forms of chronic virus production rates (left) and productive bacteria growth rates (right). With no system stress (no antibiotics), the chronic phage production rate is $\beta_C$ and the bacterial growth rate is $r_P$. As stress increases, the phage production rate saturates to $\beta_{\mathrm{max}}$ and the bacterial growth rate approaches 0.} \label{fig:beta}
\end{figure}

Finally, studies show that productive bacteria stressed by sublethal antibiotics increase viral production while decreasing cell reproduction \cite{hagens2006augmentation, secor2015filamentous}. We incorporate the increased viral production with a Hollings-like functional response; when there is no antibiotic stress in the system ($a=0$), viruses are produced at a rate $\beta_C$, and as the concentration of antibiotics increases, the viral production rate saturates at $\beta_{\mathrm{max}}$ (see Figure \ref{fig:beta}a). The functional form of the stress-dependent production rate is 
 \begin{align}
	\beta_C(a) = \beta_C + \frac{a}{h_{\beta}+a} (\beta_{\mathrm{max}}-\beta_C),
\end{align}
where $a$ is the time-dependent antibiotic concentration, $h_b$ is the concentration at which the production rate is halfway between the minimum and maximum, and $\beta_{max}$ is the maximum production rate when stress is maximum.  

We also incorporate the decreased cell reproduction when the concentration of antibiotics increases using a Hollings-like functional response. When there is no stress to the system, the cellular reproduction rate is $r_p$. As the amount of stress (antibiotics) increases, the reproduction rate approaches 0 (See Figure \ref{fig:beta}b). The functional form of the stress-dependent cellular reproduction rate is
\begin{align}
	r_P(a) =  r_P \left(1-\frac{a}{h_{r} + a}\right),
\end{align}
where $a$ is the time-dependent antibiotic concentration, and $h_r$ is the concentration at which the reproduction rate is halfway between the minimum and maximum.

The dynamical system capturing the described model behavior is
\begin{align}
& \dot{S} = \underbrace{S \left( 1 - N \right)}_{\text{growth}} - \underbrace{\eta_T S V_T - \eta_C S V_C}_{\text{infection}} + \underbrace{\gamma_T L_T + \gamma_I I_T+ \gamma_P P_C + \gamma_C L_C}_{\text{recovery}} - \underbrace{\kappa a(t)S}_{\substack{\text{antibiotic} \\ \text{death}}}  \label{eq:sus} \\
& \dot{I}_T = \underbrace{\eta_T (1-f_T) S V_T}_{\text{infection}} - \underbrace{\delta I_T}_{\text{lysis}} - \underbrace{\gamma_I I_T}_{\text{recovery}} + \underbrace{\epsilon a(t)L_T}_{\text{induction}} - \underbrace{\kappa a(t)I_T}_{\substack{\text{antibiotic} \\ \text{death}}}
\label{eq:lyt} \\
& \dot{L}_T =\underbrace{r_T L_T \left( 1-N \right)}_{\text{growth}} +  \underbrace{\eta_T f_T S V_T}_{\text{infection}} - \underbrace{\gamma_T L_T}_{\text{recovery}}-\underbrace{\epsilon a(t)L_T}_{\text{induction}}  - \underbrace{\kappa a(t)L_T}_{\substack{\text{antibiotic} \\ \text{death}}}
\label{eq:lytlat} \\
& \dot{P}_C = \underbrace{r_P(a) P_C \left( 1-N \right)}_{\text{growth}} + \underbrace{(1-f_C) \eta_C S V_C}_{\text{infection}} - \underbrace{\gamma_P P_C}_{\text{recovery}} + \underbrace{\epsilon a(t)L_C}_{\text{induction}}- \underbrace{\kappa a(t)P_C}_{\substack{\text{antibiotic} \\ \text{death}}}  
\label{eq:chron} \\
& \dot{L}_C =  \underbrace{r_C L_C \left( 1 - N \right)}_{\text{growth}} + \underbrace{f_C \eta_C S V_C}_{\text{infection}} -\underbrace{\gamma_C L_C}_{\text{recovery}} -\underbrace{\epsilon a(t)L_C}_{\text{induction}}-\underbrace{\kappa a(t)L_C}_{\substack{\text{antibiotic} \\ \text{death}}}  
\label{eq:chronlat}\\
& \dot{V}_T = \underbrace{\beta_T \delta I_T}_{\text{burst}} - \underbrace{\eta_T S V_T}_{\text{adsorption}} - \underbrace{\mu_T V_T}_{\text{degradation}}
\label{eq:lytvir} \\
& \dot{V}_C = \underbrace{\beta_C(a) P_C}_{\text{production}} - \underbrace{\eta_C S V_C}_{\text{adsorption}} - \underbrace{\mu_C V_C}_{\text{degradation}}
\label{eq:chronvir}
\end{align}

All model variables are listed in Table \ref{tab:variable} and all model parameters are listed in Table \ref{tab:param}.

	\begin{table}[!ht]
	\caption{Description of model variables in bacteria-virus system (\ref{eq:sus}-\ref{eq:chronvir}). In the system, all population densities are rescaled to be proportional to the carrying capacity and the timescale is rescaled such that all rates are relative to the susceptible bacteria's growth rate. See supplemental materials for rescaling process.}
	\renewcommand{\arraystretch}{1.5}
\footnotesize
	\begin{tabular}{| c  p{9.5cm} p{1.3cm} | }  \hline 
	{\bf Variable} & {\bf Meaning} & {\bf Units} \\  \hline 
	$S$                  & susceptible bacteria as a proportion of carrying capacity & unitless  \\
	$I_T$               & lytic bacteria as a proportion of carrying capacity & unitless \\
	$L_T$              & latent lytic bacteria as a proportion of carrying capacity  & unitless   \\
	$P_C$             & productive bacteria as a proportion of carrying capacity & unitless \\
	$L_C$              & latent chronic bacteria as a proportion of carrying capacity & unitless \\
	$N$                  & all bacteria ($S + I_T +L_T +P_C + L_C$) & unitless \\
	$V_T$              & ratio of free temperate viruses to bacteria carrying capacity & $\displaystyle\frac{\text{PFU}}{\text{CFU}}$ \\
	$V_C$              & ratio of free chronic viruses to bacteria carrying capacity & $\displaystyle\frac{\text{PFU}}{\text{CFU}}$  \\
	$t$                    & time rescaled by intrinsic growth rate & unitless \\
	\hline
	\end{tabular}
	\label{tab:variable}
	\end{table}

	\begin{table}[h!]
	\caption{Description of model parameters in bacteria-virus system (\ref{eq:sus}-\ref{eq:chronvir}). See supplemental materials for parameter selection process and model rescaling process.}
\footnotesize
	\begin{tabular}{| c  p{8.5cm} p{1.8cm}  p{1.5cm} p{2.2cm} |}  \hline 
	{\bf Parameter} & {\bf Meaning} & {\bf Range$^a$} & {\bf Baseline} & {\bf Sources} \\  \hline 
	$r_T, r_C$ & growth rates of (respectively) latent lytic and latent chronic bacteria, relative to the susceptible bacteria growth rate$^b$ & [0.5, 3]$^c$ & 1 & \cite{shapiro2016evolution} \\ 
	$r_P(a)$ & growth rate of productive bacteria, a decreasing function of the antibiotic concentration $a$ & [0.5, 3]$^c$ & 1 & \cite{shapiro2016evolution} \\ 
	$\eta_T, \eta_C$ & infection rate of (respectively) temperate and chronic viruses  & [0.38, 14.7]$^e$  & 1  & \cite{sinha2017silico} \\ 
	$\gamma_T, \gamma_P, \gamma_C$ & recovery rates of (respectively) latent lytic, productive, and latent chronic bacteria  & [0,1000]$^f$  & 0.67$^g$  & \cite{brussow2004phages,horvath2010crispr} \\
	$\gamma_I$ & recovery rates of lytic bacteria  & [0,1000]  & 0$^h$  & \cite{brussow2004phages,horvath2010crispr} \\ 
	$\delta$ & rate at which lytic infection leads to bursting &  [1.5, 7.8]$^i$ & 4 & \cite{yu2017characterization,el2015isolation} \\ 
	$f_T$ & lysogen frequency for temperate viruses & [0, 0.9]  & 0.01  & \cite{calendar2006bacteriophages,oppenheim2007new,volkova2014modeling} \\ 
	$f_C$ & lysogen frequency for chronic viruses & [0, 0.9]  & 0$^g$  & \cite{calendar2006bacteriophages,oppenheim2007new,volkova2014modeling} \\ 
	$\beta_T$ & burst size for bacteria infected with $V_T$ &  [10, 1000] & 100 & \cite{yu2017characterization,el2015isolation,latino2014novel,schrader1997bacteriophage,ceyssens2010molecular,garbe2011sequencing,you2002effects}  \\ 
	$\beta_C(a)$ & viral production rate for bacteria infected with $V_C$, an increasing function of antibiotic concentration $a$ & [5, 200] & 20 & \cite{clifton2019modeling}   \\
	$\mu_T, \mu_C$ & degradation rate of (respectively) free temperate viruses and free chronic viruses &  [0.9, 3.6]$^j$ & 1  & \cite{heldal1991production}  \\ 
    $\kappa$  & death rate per concentration of antibiotics & [0, 3.5] & 1 & \cite{spalding2018mathematical, grillon2016comparative} \\
    $\epsilon$  & lysis rate per concentration of antibiotics & [0, 2] & 1 & \cite{spalding2018mathematical, grillon2016comparative} \\
    $A$ & maximum concentration of one standard antibiotic dose ($\mu$g/mL) & [0, 10]$^k$ & 1 & \cite{grillon2016, fong1986} \\
    $k$ & metabolic decay rate of antibiotic within the system & [1e-3,0.6]$^l$ & 0.3 & \cite{spalding2018mathematical,zhanel2006review,wingender1984pharmacokinetics} \\
    $h_{\beta}, h_r$ & stress level at which growth rate is half the maximum &  & 1 & \\
	\hline
	\end{tabular}
	 \begin{flushleft} \footnotesize
	$^a$all parameter ranges are taken for the human pathogens \textit{P.~aeruginosa} or \textit{E.~coli} and their viruses, unless otherwise noted.
	 $^b$growth rate is approximately 5.1e-3 min$^{-1}$ for \textit{P.~aeruginosa} grown \textit{in vitro}, but is highly variable in human hosts.
	 $^c$estimates based on \textit{E.~coli} and M13 phage. 
	 $^d$stable bacterial density in human hosts is highly variable; a study of viable \textit{P.~aeruginosa} densities in sputum of 12 patients with cystic fibrosis not undergoing treatment ranged from 5.3e3 CFU/mL to 1.8e11 CFU/mL.  
	 $^e$estimates based on \textit{E.~coli} and $\lambda$ phage. 
	$^f$a wide range of recovery rates has been found; some proviruses are viable over evolutionary timescales and some proviruses are inactivated nearly instantly by CRISPR systems.
	$^g$estimated from viral steady state density (see Results section).
	  $^h$selected to be 0 to simplify model analysis; allowing $\gamma_I = \gamma_T$ produces qualitatively similar results, so the increased model complexity is not justified.
	  $^i$low estimate is for PAXYB1 phage and PAO1 host, high estimate is for $\varphi$PSZ1 phage and PAO1 host.
	  $^j$low estimate is for viruses extracted from Raunefjorden, high estimate is for viruses extracted from Bergen Harbor (strains unknown). $^k$blood serum concentration of ciprofloxacin and levofloxacin. $^l$antibiotic is levofloxacin.
	  \end{flushleft}
	\label{tab:param}
	\end{table}

\section{Results}
Because patients infected with dangerous bacterial infections, particularly \textit{P.~aeruginosa}, are typically treated with antibiotics at the time of bacterial detection \cite{hoiby2017diagnosis,hoiby2015escmid}, we investigate the effects of antibiotic administration on the bacteria-phage ecosystem. We assume that the antibiotics trigger an SOS response in the bacterium which induces the production of latent temperate and chronic viruses \cite{fothergill2011effect, ptashne1986genetic, nanda2015impact}. Not all antibiotics induce phage via the SOS response \cite{fothergill2011effect}, so we focus only on the types of antibiotics known to do so (e.g., quinolones like levofloxacin and ciprofloxacin) \cite{zhang2000quinolone,comeau2007phage}. We assume that even antibiotic-resistant bacteria induce viruses in the presence of antibiotics, which has been demonstrated for several classes of antibiotics \cite{redgrave2014fluoroquinolone,valencia2017ciprofloxacin,brazas2005ciprofloxacin,james2012differential,fothergill2011effect}. 

We also assume that as the concentration of antibiotics increases, the production rate of the chronic viruses increases and the reproduction rate of productive bacteria decreases \cite{hagens2006augmentation, secor2015filamentous}. This assumption is critical to the main results presented here, so we thoroughly explore the effect of chronic virus production as a function of antibiotic administration frequency in the Sensitivity Analysis section.

\begin{figure}[htbp]
  \centering
    \includegraphics[width=\textwidth]{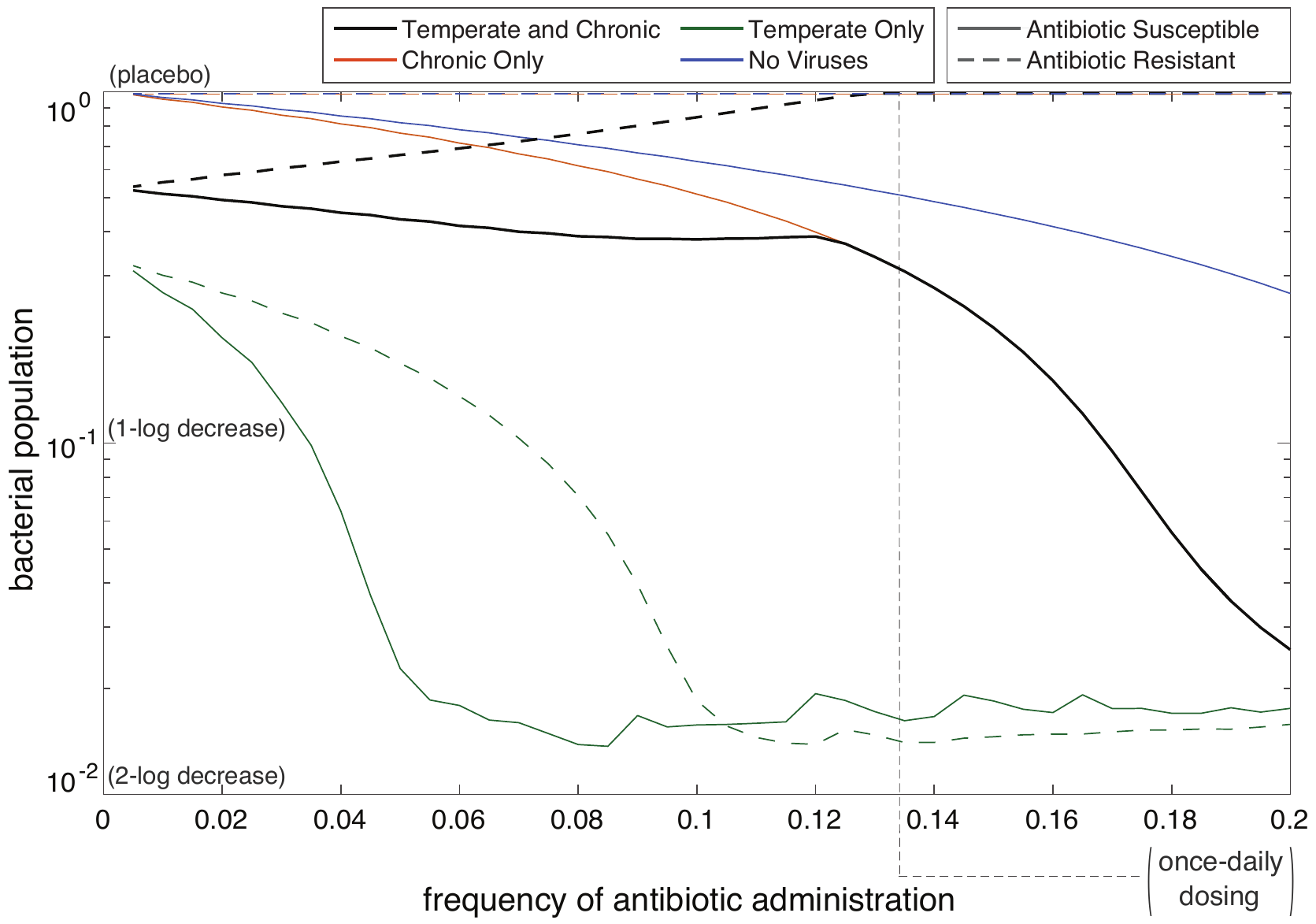}
      \caption{Simulation of the effect of antibiotic administration frequency on the time-averaged bacterial population when infected with various bacteriophages in the human host. Time is scaled by the growth rate of \textit{P.~aeruginosa} ($5.1$e$-3$ min$^{-1}$), so 0.136 in unitless frequency units is approximately one dose per 24 hours. In the system, all population densities are rescaled to be proportional to the carrying capacity. Thus, the ``placebo'' condition (no antibiotic treatment) results in the bacterial population reaching carrying capacity (1 in unitless density).} \label{fig:results}
\end{figure}

Figure \ref{fig:results} shows the time-averaged bacterial population level as a function of the frequency of antibiotic administration. Antibiotic susceptible bacterial populations are represented with solid lines, and antibiotic resistant bacterial populations are represented with dashed lines. 

\subsection{Antibiotic Susceptible}
Figure \ref{fig:results} illustrates that no matter the antibiotic frequency, the bacterial population is largest when not infected with viruses. When the SOS response is triggered in bacteria without bacteriophages, no lysis from rapid bacteriophage reproduction occurs. Instead, the bacteria can only die from antibiotics. Alternatively, when bacteriophages are present, the bacterial population decreases at a faster rate. When the frequency of antibiotics is below 0.12, the presence of temperate and chronic viruses has a larger negative impact on the bacterial population than when only chronic viruses are present. However, once the antibiotic frequency is larger than 0.12, the bacterial populations become equivalent when temperate and chronic and only chronic are present. When no antibiotics are present, the chronic bacterial population reaches carrying capacity because the bacteria are not lysing or being killed by antibiotics. However, when chronic and temperate viruses are present with no antibiotics, the temperate viruses can still lyse the bacteria. This accounts for the difference in bacterial population when frequency of antibiotic administration is zero. The two populations converge when the antibiotic frequency is greater than 0.12 because chronic viruses out-compete temperate viruses due to the stress-induced chronic virus production rate increase. 
 
Antibiotics have the largest effect on the bacterial population infected with only temperate viruses. When infected with temperate viruses, the bacterial population severely decreases as frequency of antibiotics increases because latent bacteria are being induced by antibiotics into the lytic lifestyle. The apparent noise in population levels when the frequency is greater than 0.08 is likely due to chaos in the system. Studies show that chaos is commonly seen in periodically forced predator-prey models \cite{tang2003,taylor2013}. Further exploration of this phenomenon is beyond the scope of this paper.  

\subsection{Antibiotic Resistant}
When no bacteriophages are present in the system or when only chronic viruses are present in the system, the antibiotic-resistant bacterial populations remain at carrying capacity as the frequency of antibiotics increases. The bacterial population reaches carrying capacity when no viruses are present because there exists no mechanism (antibiotic or virus) to kill the bacteria. The bacterial population also reaches carrying capacity when only chronic viruses are present because chronic viruses do not kill their host in order to reproduce.

The behaviors of the antibiotic-resistant bacterial population change when both viruses are in the system and when only temperate viruses are in the system. The bacterial population infected by only temperate viruses decreases as the frequency of antibiotics is increased. Decreasing at an increasing rate, we see that the bacterial population levels off at 0.11. This is because as the frequency of antibiotics increases, more latent temperate viruses are induced into the lytic cycle. The dynamics of the antibiotic resistant bacteria infected with temperate bacteriophages are similar to when the bacteria are antibiotic susceptible, however when the bacteria are antibiotic resistant, it requires a higher frequency of antibiotics to significantly decrease the bacterial population. 

Contrary to expectations, as antibiotic frequency increases, the antibiotic-resistant bacterial population infected with temperate and chronic viruses increases. At a frequency of roughly 0.13, the bacterial population infected with temperate and chronic viruses reaches carrying capacity. While antibiotics are not killing the antibiotic resistant bacteria through the intended mechanisms, they cause the induction of the SOS response in bacteria. This induction causes the production of chronic viruses and lysis of temperate infected bacteria. At higher antibiotic frequencies, all of the temperate infected bacteria lyse. The only bacteria that are left are the bacteria that have been infected by chronic viruses. Chronic viruses, having out-competed temperate viruses (a natural predator of bacteria), leave bacteria invulnerable to control. This phenomenon has clinical implications in that, contrary to the intended outcome, antibiotics can counterintuitively increase the bacterial population.

\subsection{Sensitivity Analysis} \label{sens}
\begin{figure}[htbp]
  \centering
    \includegraphics[width=0.75\textwidth]{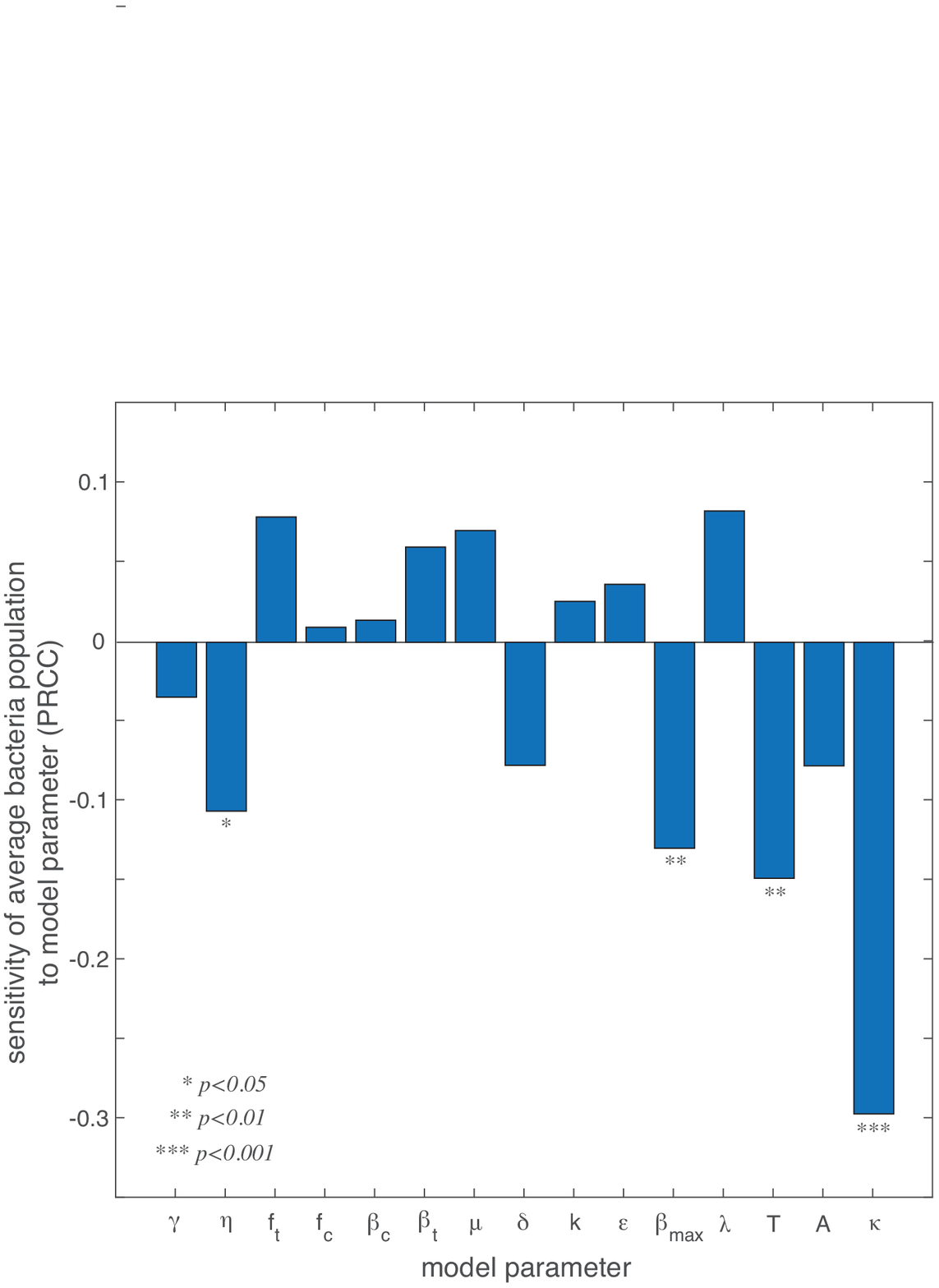}
      \caption{Sensitivity of parameters on average bacteria population. The sensitivity analyses uses Latin hypercube sampling (LHS) of parameter space and partial rank correlation coefficients (PRCC) \cite{marino2008methodology}. All parameter values are taken near the baselines in Table \ref{tab:param}. Initially, $S(0)=1e-3, V_T(0)=V_C(0)=1e-7$. 500 simulations were completed. Note that asterisks indicate significance. See the supplemental material for technical details.} \label{fig:sensitivity}
\end{figure}

While we have used \textit{P.~aeruginosa} infections as a case study, this model is theoretically applicable to any ecosystem with bacteria, temperate viruses, and chronic viruses. Therefore, we conduct a sensitivity and uncertainty analysis to understand the impact of each model parameter on the clinical outcome of interest: bacteria population size.
Because the relationships between the parameters and the bacterial population size are generally nonlinear but monotonic, we use Latin Hypercube Sampling (LHS) of parameter space and partial rank correlation coefficients (PRCC). Following Marino et al.~\cite{marino2008methodology}, the sensitivity analysis indicates that the parameters $\kappa$, $T$, $\beta_{max}$, and $\eta$ have a significant ($p<0.05$) impact on the bacterial population (Figure \ref{fig:sensitivity}). The antibiotic death rate ($\kappa$), has the largest magnitude of PRCC, indicating that a small increase in $\kappa$ will result in the largest decrease of the bacterial population. The period of antibiotic administration ($T$), the infection rate $\eta$, and the maximum chronic virus production rate $\beta_{max}$ also have a negative correlation with the bacterial population size.

\begin{figure}[htbp]
  \centering
    \includegraphics[width=0.8\textwidth]{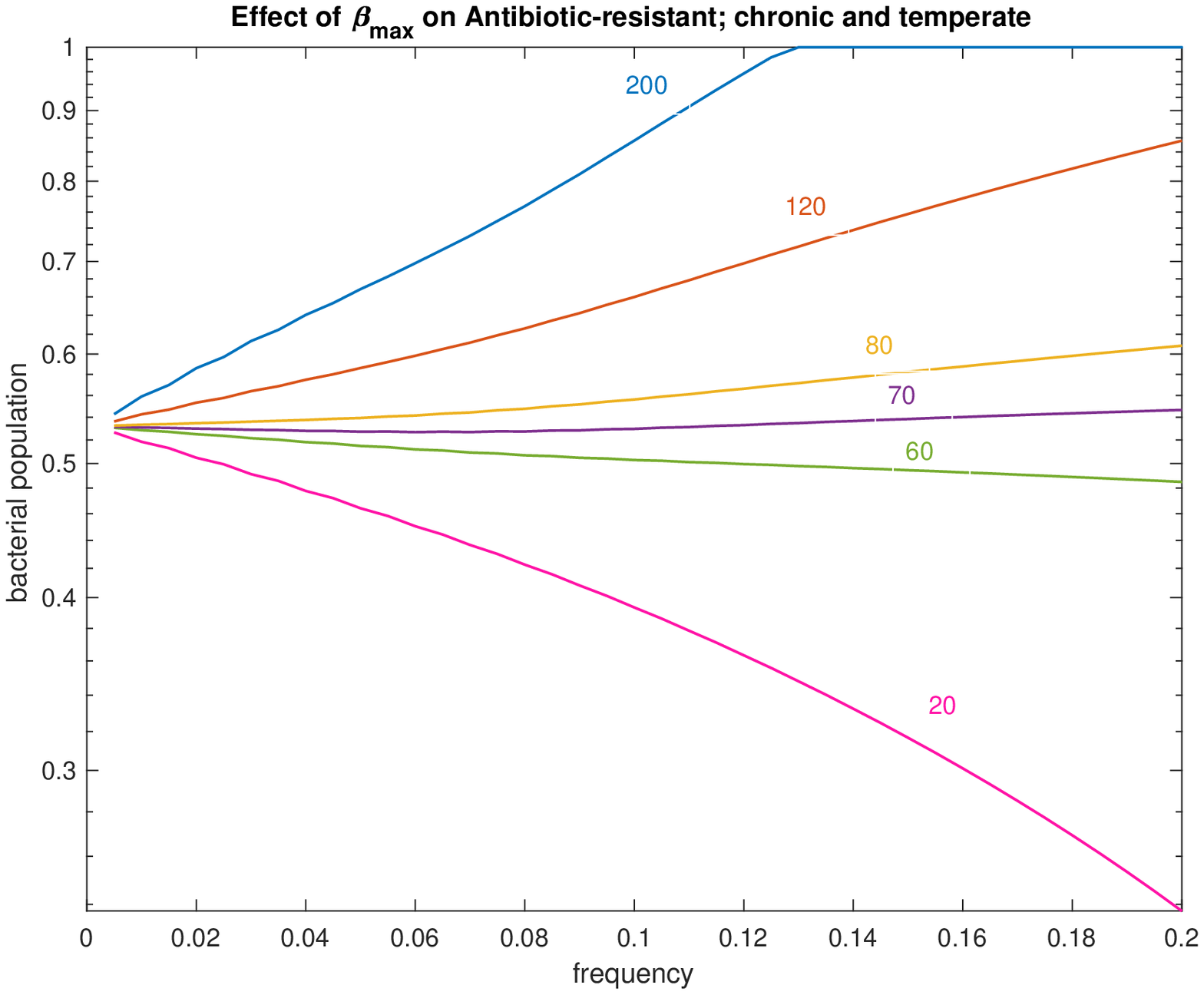}
      \caption{Impact of $\beta_{max} $ on resistant bacterial population when temperate and chronic bacteriophages are present. As $\beta_{max}$ decreases, the effect of antibiotic frequency on bacterial population flips from a positive slope to a negative slope.} \label{fig:beta_max}
\end{figure}

For any particular bacteria-phage-antibiotic system, the parameters $\kappa, T$, and $\eta$ should be well-characterised. It is less likely that $\beta_{max}$ will be well-characterized because chronic viruses have not been studied as extensively as lytic and temperate viruses. 
We wish to understand the impact of $\beta_{max}$ on the bacterial population when the population is antibiotic resistant and infected with both temperate and chronic viruses; this scenario is the most common for a patient infected with $\textit{P.~aeruginosa}$ \cite{james2015lytic,burgener2019,geller2011levofloxacin}. Figure \ref{fig:beta_max} shows that as $\beta_{max}$ decreases, the slope of bacterial population as a function frequency of antibiotic decreases until it ultimately becomes negative around $\beta_{max} = 70$. This shows that virus competition is only detrimental to infection control if chronic virus production is sufficiently increased by antibiotic stress.

\section{Discussion}
The implications of our model can theoretically be applied to any bacterial infection with naturally present temperate and chronic viruses, but the results presented in this paper focus on treatment of \textit{P. aeruginosa} infections in patients with cystic fibrosis. Clinical trials of antibiotics confirm that both temperate and chronic viral lifestyles are present in the lungs of most patients with cystic fibrosis \cite{james2015lytic,burgener2019}, and around 60\% of \textit{P. aeruginosa} in sputum samples are antibiotic resistant \cite{geller2011levofloxacin}. Despite the rise of antibiotic-resistant bacterial infections, antibiotics remain a common treatment for patients with cystic fibrosis \cite{proesmans2013}. Unfortunately, our model suggests that antibiotic treatment may be counterproductive under the most likely treatment conditions. These results highlight the need for personalized medicine approaches for treating antibiotic-resistant bacterial infection which respect, or even exploit, the underlying ecology within the patient.

\subsection{Limitations and Future Steps}
Our results are limited by assumptions made in order to create a simple, analytically tractable model. First, the model assumes that the antibiotics present induce viral production through the SOS response. Not all antibiotics cause an SOS response in bacteria. Thus, the model is applicable to antibiotics known to trigger an SOS response (e.g., quinolone antibiotics like levofloxacin and ciprofloxacin) \cite{fothergill2011effect}. This is not a major limitation as treatment of \textit{P. aeruginosa} infections often include quinolone antibiotics \cite{geller2011levofloxacin, hodson1987oral}.

The model also assumes mass action infection of bacteria by phage. Due to the biological nature of bacteria, this assumption breaks down at large bacterial densities. Bacteria like \textit{P. aeruginosa} are known to form biofilms at high densities, which effectively saturates the infection rate \cite{harper2014bacteriophages}. Previous studies have used a Hollings Type II functional response to model the infection rate, and the results are not qualitatively different \cite{clifton2019modeling}. However, a more sophisticated model would include biofilm formation in the infection assumptions.  

The phenomenon of multiple infections, where more than one phage infects a single bacterium at one time is excluded from our model.  This is not a significant limitation because when infected with bacteriophages, \textit{P. aeruginosa} produces superinfection exclusion proteins which prevent multiple infections by the same phage type \cite{james2012differential, heo2007genome}. Another limitation of our model is that we do not consider the severity of an infection. The concept of multiplicity, where more viruses infecting the host may increase the severity of the illness, is not considered because the severity of infection is not included in our model. 

From the sensitivity analysis, we know that parameters $\kappa$, $T$, $\beta_{max}$, and $\eta$ have the largest impact on the accuracy of our model. While $T$ is determined by the clinician and $\kappa$ and $\eta$ are generally well-characterized for a broad range of phages and antibiotics, $\beta_{max}$ has not been definitively determined experimentally. Figure \ref{fig:beta_max} illustrates the impact of $\beta_{max}$ on antibiotic resistant bacteria infected with temperate and chronic viruses. Figure \ref{fig:beta_max} illustrates the importance of having a precise measurement of $\beta_{max}$. Further experiments are needed to determine a precise value of $\beta_{max}$, which would allow the model to be more readily applicable in a clinical setting. 

Finally, because our model does not incorporate evolutionary dynamics, such as the development of antibiotic or phage resistance, our results are only applicable on short time scales. Our model could serve as the base of a more sophisticated model that includes both ecological and evolutionary dynamics.

\section{Conclusion}
This model demonstrates the potential for an antagonistic effect of virus competition on antibiotic treatment of resistant bacterial infections. Both temperate and chronic virus lifestyles are naturally present in human hosts, and antibiotic resistance is a growing concern. Because certain combinations of phages and antibiotics can exhibit a synergistic effect when treating bacterial infections, this work suggests that a more personalized medicine approach may provide better clinical outcomes. Treatment plans tailored to the underlying bacteria-phage ecology naturally present within human hosts are critical to controlling bacterial infections, especially when the bacteria are resistant to antibiotics.

\section{Acknowledgements}
This work was funded in part by the National Science Foundation grant DMS-1815764 (ZR), the Cystic Fibrosis Foundation grant WHITAK16PO (RJW), and an Allen Distinguished Investigator Award (RJW). The funders had no role in study design, data collection and analysis, decision to publish, or preparation of the manuscript.

The authors thank Jayadevi H.~Chandrashekhar and George A.~O'Toole for discussions that informed biological aspects of this work. Thanks are also due to Ted Kim and Karna Gowda for assistance with parameter selection. The authors also thank Laura Suttenfield and Alan Collins for suggestions that improved the clarity of the paper.

\section{Data availability}  
All software (Matlab .m files) will be made publicly available via the Illinois Data Bank.

\section{Competing interests}
The authors declare no competing interests.

\bibliographystyle{ieeetr}
\bibliography{library}

\newpage
\section{Appendix}
\subsection{Model rescaling}  \label{non}
The dimensional model system (\ref{eq:sus}-\ref{eq:chronvir}) below can be rescaled so that time and bacterial density are unitless quantities. Although the viral density can also be nondimensionalized to be unitless, we have elected (for the sake of interpretability) to rescale viral density to have units of PFU/CFU. 
\begin{align}
& \dot{\tilde{S}} = \underbrace{\tilde{r}_S \tilde{S} \left( 1-\frac{\tilde{N}}{\tilde{K}} \right)}_{\text{growth}} - \underbrace{\tilde{\eta}_T \tilde{S} \tilde{V}_T - \tilde{\eta}_C \tilde{S} \tilde{V}_C}_{\text{infection}} + \underbrace{\tilde{\gamma}_T \tilde{L}_T + \tilde{\gamma}_I \tilde{I}_T+ \tilde{\gamma}_P \tilde{P}_C + \tilde{\gamma}_C \tilde{L}_C}_{\text{recovery}} - \underbrace{\tilde{\kappa} \tilde{a}(\tilde{t}) \tilde{S}}_{\substack{\text{antibiotic} \\ \text{death}}} 
\label{eq:susd} \\
& \dot{\tilde{I}}_T = \underbrace{\tilde{\eta}_T (1-\tilde{f}_T) \tilde{S} \tilde{V}_T}_{\text{infection}} - \underbrace{\tilde{\delta} \tilde{I}_T}_{\text{lysis}} - \underbrace{\tilde{\gamma}_I \tilde{I}_T}_{\text{recovery}} + \underbrace{\tilde{\epsilon} \tilde{a}(\tilde{t})\tilde{L_T}}_{\text{induction}} - \underbrace{\tilde{\kappa} \tilde{a}(\tilde{t})\tilde{I_T}}_{\substack{\text{antibiotic} \\ \text{death}}}
\label{eq:lytd} \\
& \dot{\tilde{L}}_T = \underbrace{\tilde{r}_T \tilde{L}_T \left( 1-\frac{\tilde{N}}{\tilde{K}} \right)}_{\text{growth}} + \underbrace{\tilde{\eta}_T \tilde{f}_T \tilde{S} \tilde{V}_T}_{\text{infection}} - \underbrace{\tilde{\gamma}_T \tilde{L}_T}_{\text{recovery}} -\underbrace{\tilde{\epsilon} \tilde{a}(\tilde{t})\tilde{L_T}}_{\text{induction}}  - \underbrace{\tilde{\kappa} \tilde{a}(\tilde{t})\tilde{L_T}}_{\substack{\text{antibiotic} \\ \text{death}}}
\label{eq:lytlatd} \\
& \dot{\tilde{P}}_C = \underbrace{\tilde{r}_P \tilde{P}_C \left( 1-\frac{\tilde{N}}{\tilde{K}} \right)}_{\text{growth}} + \underbrace{(1-\tilde{f}_C) \tilde{\eta}_C \tilde{S} \tilde{V}_C}_{\text{infection}} - \underbrace{\tilde{\gamma}_P \tilde{P}_C}_{\text{recovery}} + \underbrace{\tilde{\epsilon} \tilde{a}(\tilde{t})\tilde{L_C}}_{\text{induction}}- \underbrace{\tilde{\kappa} \tilde{a}(\tilde{t})\tilde{P_C}}_{\substack{\text{antibiotic} \\ \text{death}}}  
\label{eq:chrond} \\
& \dot{\tilde{L}}_C =  \underbrace{\tilde{r}_C \tilde{L}_C \left( 1-\frac{\tilde{N}}{\tilde{K}} \right)}_{\text{growth}} + \underbrace{\tilde{f}_C \tilde{\eta}_C \tilde{S} \tilde{V}_C}_{\text{infection}} -\underbrace{\tilde{\gamma}_C \tilde{L}_C}_{\text{recovery}} -\underbrace{\tilde{\epsilon} \tilde{a}(\tilde{t})\tilde{L_C}}_{\text{induction}}-\underbrace{\tilde{\kappa} \tilde{a}(\tilde{t})\tilde{L_C}}_{\substack{\text{antibiotic} \\ \text{death}}}  
\label{eq:chronlatd}\\
& \dot{\tilde{V}}_T = \underbrace{\tilde{\beta}_T \tilde{\delta} \tilde{I}_T}_{\text{burst}} - \underbrace{\tilde{\eta}_T \tilde{S} \tilde{V}_T}_{\text{adsorption}} - \underbrace{\tilde{\mu}_T \tilde{V}_T}_{\text{degradation}}
\label{eq:lytvird} \\
& \dot{\tilde{V}}_C = \underbrace{\tilde{\beta}_C \tilde{P}_C}_{\text{production}} - \underbrace{\tilde{\eta}_C \tilde{S} \tilde{V}_C}_{\text{adsorption}} - \underbrace{\tilde{\mu}_C \tilde{V}_C}_{\text{degradation}}
\label{eq:chronvird}
\end{align}

Suppose equations (\ref{eq:susd}-\ref{eq:chronvird}) have bacterial density units of CFU/mL, viral density units of PFU/mL, and time units of minutes; then equations (\ref{eq:susd}-\ref{eq:chronlatd}) have units of CFU/mL/min, and equations (\ref{eq:lytvird}-\ref{eq:chronvird}) have units of PFU/mL/min. We will make the following substitutions into the system:
\begin{align*}
\tilde{S} &= \tilde{K} S &
\tilde{I}_T &= \tilde{K} I_T \\
\tilde{L}_T &= \tilde{K} L_T &
\tilde{P}_C &= \tilde{K} P_C \\
\tilde{L}_C &= \tilde{K} L_C &
\tilde{N} &= \tilde{K} N \\
\tilde{V}_T &= \tilde{K} V_T &
\tilde{V}_C &= \tilde{K} V_C \\
\tilde{t} &= \frac{t}{\tilde{r}_s} &
\end{align*}
where $\displaystyle S, \dots , N$ are unitless bacterial densities, $\displaystyle V_T,  V_C$ have units of PFU/CFU (a virus to bacteria ratio), and $t$ is a unitless time. An alternate substitution of $\displaystyle \tilde{V}_T = \tilde{\beta}_T \tilde{K} V_T$ and $\displaystyle \tilde{V}_C = \tilde{\beta}_T \tilde{K} V_C$ would have led to a nondimensionalization of the system, but the nondimensionalized viral densities would have been more challenging to interpret; therefore we eschew this option. 

We will illustrate the rescaling process for bacterial dynamics with equation (\ref{eq:susd}), and equations (\ref{eq:lytd}-\ref{eq:chronlatd}) are similar:
\begin{equation*}
\frac{\mathrm{d} \tilde{S}}{\mathrm{d} \tilde{t}} =  \underbrace{\tilde{r}_S \tilde{S} \left( 1-\frac{\tilde{N}}{\tilde{K}} \right)}_{\text{growth}} - \underbrace{\tilde{\eta}_T \tilde{S} \tilde{V}_T - \tilde{\eta}_C \tilde{S} \tilde{V}_C}_{\text{infection}} + \underbrace{\tilde{\gamma}_T \tilde{L}_T + \tilde{\gamma}_I \tilde{I}_T+ \tilde{\gamma}_P \tilde{P}_C + \tilde{\gamma}_C \tilde{L}_C}_{\text{recovery}}  -\underbrace{\tilde{\kappa} \tilde{a}(\tilde{t})\tilde{S}}_{\text{antibiotic death} \\}
\end{equation*}
After making the substitutions into equation (\ref{eq:susd}), we get
\begin{align*}
\frac{\mathrm{d} (\tilde{K} {S})}{\mathrm{d} ({t}/\tilde{r}_S)} =& \underbrace{\tilde{r}_S \tilde{K} {S} \left( 1-\frac{\tilde{K} {N}}{\tilde{K}} \right)}_{\text{growth}} - \underbrace{\tilde{\eta}_T \tilde{K} {S} \tilde{K} {V}_T - \tilde{\eta}_C \tilde{K} {S} \tilde{K} {V}_C}_{\text{infection}}\\
& + \underbrace{\tilde{\gamma}_T \tilde{K} {L}_T +\tilde{\gamma}_I \tilde{K} {I}_T+ \tilde{\gamma}_P \tilde{K} {P}_C + \tilde{\gamma}_C \tilde{K} {L}_C}_{\text{recovery}}- \underbrace{\tilde{\kappa} \tilde{a}(\tilde{t})\tilde{K}S}_{\text{antibiotic death}}
\end{align*}
We now divide the equation by $\tilde{K} \tilde{r}_S$:
\begin{equation*}
\frac{\mathrm{d} {S}}{\mathrm{d} {t}} = \underbrace{{S} \left(1-{N} \right)}_{\text{growth}} - \underbrace{\frac{\tilde{\eta}_T \tilde{K}}{\tilde{r}_S}{S}{V}_T - \frac{\tilde{\eta}_C \tilde{K}}{\tilde{r}_S} {S} {V}_C}_{\text{infection}} + \underbrace{\frac{\tilde{\gamma}_T}{\tilde{r}_S} {L}_T + \frac{\tilde{\gamma}_I}{\tilde{r}_S} {I}_T+ \frac{\tilde{\gamma}_P}{\tilde{r}_S} {P}_C + \frac{\tilde{\gamma}_C}{\tilde{r}_S}  {L}_C}_{\text{recovery}}- {\underbrace{\frac{\tilde{\kappa} \tilde{a}(\tilde{t}) }{\tilde{r_s}} S}_{\text{antibiotic death}}}
\end{equation*}
The transformed parameters for growth, infection, and recovery are now evident:
\begin{align*}
{\eta}_T &= \frac{\tilde{\eta}_T \tilde{K}}{\tilde{r}_S} &
{\eta}_C &= \frac{\tilde{\eta}_C \tilde{K}}{\tilde{r}_S} \\
{\gamma}_T &= \frac{\tilde{\gamma}_T}{\tilde{r}_S} &
{\gamma}_I &= \frac{\tilde{\gamma}_I}{\tilde{r}_S} \\
{\gamma}_P &= \frac{\tilde{\gamma}_P}{\tilde{r}_S} &
{\gamma}_C &= \frac{\tilde{\gamma}_C}{\tilde{r}_S}
\end{align*}
The transformed parameters for antibiotic death require more manipulation. Because 
\[\tilde{a}(\tilde{t}) = \tilde{A} \, \sum_{i=1}^N H(\tilde{t}-\tilde{t}_i) \, e^{-\tilde{k} (\tilde{t}-\tilde{t}_i)}, \]
we have the unitless parameter grouping\footnote{The argument of the Heaviside function can be multiplied by any positive number without changing the output.}
\begin{align*} \frac{\tilde{\kappa} \tilde{a}(\tilde{t})}{\tilde{r_s}} &= \frac{\tilde{\kappa} }{\tilde{r_s}} \tilde{A} \, \sum_{i=1}^N H(\tilde{r}_S\tilde{t}-\tilde{r}_S\tilde{t}_i) \, e^{-\frac{\tilde{k}}{\tilde{r}_S} (\tilde{r}_S\tilde{t}-\tilde{r}_S\tilde{t}_i)} \\
&= \kappa \, A \, \sum_{i=1}^N H(t-t_i) \, e^{-k(t-t_i)} \\
&= \kappa \, a(t), 
\end{align*}
where the transformed parameters are
\begin{align*}
\kappa &= \frac{\tilde{\kappa} }{\tilde{r_s}} &
A &= \tilde{A} & k &= \frac{\tilde{k} }{\tilde{r_s}}
\end{align*}
With the previous substitutions, equation (\ref{eq:susd}) reduces to
\begin{equation*}
\frac{\mathrm{d} {S}}{\mathrm{d} {t}} =  \underbrace{{S} \left( 1-{N} \right)}_{\text{growth}} - \underbrace{{\eta}_T {S} V_T - {\eta}_C {S} V_C}_{\text{infection}} + \underbrace{{\gamma}_T {L}_T +{\gamma}_I {I}_T+ {\gamma}_P {P}_C + {\gamma}_C {L}_C}_{\text{recovery}} - \underbrace{\kappa a(t)S}_{\substack{\text{antibiotic} \\ \text{death}}} 
\end{equation*}
Equations (\ref{eq:lytd}-\ref{eq:chronlatd}) are rescaled in a similar manner. Substituting the transformations into equations (\ref{eq:lytvird}-\ref{eq:chronvird}) yields
\begin{align*}
\frac{\mathrm{d} (\tilde{K} {V}_T)}{\mathrm{d} ({t}/\tilde{r}_S)} &= \underbrace{\tilde{\beta}_T \tilde{\delta} \tilde{K} {I}_T}_{\text{burst}} - \underbrace{\tilde{\eta}_T \tilde{K} {S} \tilde{K} {V}_T}_{\text{adsorption}} - \underbrace{\tilde{\mu}_T \tilde{K} {V}_T}_{\text{degradation}} \\
\frac{\mathrm{d} (\tilde{K} {V}_C)}{\mathrm{d} ({t}/\tilde{r}_S)} &= \underbrace{\tilde{\beta}_C \tilde{K} {P}_C}_{\text{production}} - \underbrace{\tilde{\eta}_C \tilde{K} {S} \tilde{K} {V}_C}_{\text{adsorption}} - \underbrace{\tilde{\mu}_C \tilde{K} {V}_C}_{\text{degradation}}
\end{align*}
Again dividing by $\tilde{K} \tilde{r}_S$, we get:
\begin{align*}
\frac{\mathrm{d} {V}_T}{\mathrm{d}{t}} &= \underbrace{\tilde{\beta}_T \frac{\tilde{\delta}}{\tilde{r}_S} {I}_T}_{\text{burst}} - \underbrace{\frac{\tilde{\eta}_T \tilde{K}}{\tilde{r}_S} {S} {V}_T}_{\text{adsorption}} - \underbrace{\frac{\tilde{\mu}_T}{\tilde{r}_S} {V}_T}_{\text{degradation}} \\
\frac{\mathrm{d} {V}_C}{\mathrm{d} {t}} &= \underbrace{\frac{\tilde{\beta}_C}{\tilde{r}_S} {P}_C}_{\text{production}} - \underbrace{\frac{\tilde{\eta}_C \tilde{K}}{\tilde{r}_S} {S} {V}_C}_{\text{adsorption}} - \underbrace{\frac{\tilde{\mu}_C}{\tilde{r}_S} {V}_C}_{\text{degradation}}
\end{align*}
Adding to our list of transformed parameters
\begin{align*}
{\beta}_T &= \tilde{\beta}_T &
{\beta}_C &= \frac{\tilde{\beta}_C}{\tilde{r}_S} \\
{\mu}_T &= \frac{\tilde{\mu}_T}{\tilde{r}_S} &
{\mu}_C &= \frac{\tilde{\mu}_C}{\tilde{r}_S} \\
{\delta} &= \frac{\tilde{\delta}}{\tilde{r}_S} &
\end{align*}
the viral density equations reduce to
\begin{align*}
\frac{\mathrm{d} {V}_T}{\mathrm{d}{t}} &= \underbrace{{\beta}_T {\delta} {I}_T}_{\text{burst}} - \underbrace{{\eta}_T {S} {V}_T}_{\text{adsorption}} - \underbrace{{\mu}_T {V}_T}_{\text{degradation}} \\
\frac{\mathrm{d} {V}_C}{\mathrm{d} {t}} &= \underbrace{{\beta}_C {P}_C}_{\text{production}} - \underbrace{{\eta}_C {S} {V}_C}_{\text{adsorption}} - \underbrace{{\mu}_C {V}_C}_{\text{degradation}}
\end{align*}

The full set of transformed parameters is listed in Table \ref{tab:paramfull}. 

	\begin{table}[!t]
	\caption{Summary of model parameter groupings after rescaling of system (\ref{eq:susd}-\ref{eq:chronvird}).}
	\begin{center} \renewcommand{\arraystretch}{1.75}
\footnotesize
	\begin{tabular}{| p{2.1cm} | p{1.6cm} | c | p{6.5cm} | p{1.5cm} |}  \hline 
	{\bf Parameter} & {\bf Units} & & {\bf Associated parameter grouping} & {\bf Units} \\  \hline 
	$\tilde{r}_S$ & min$^{-1}$ &  & $\displaystyle {r}_S = \frac{\tilde{r}_S}{\tilde{r}_S} = 1$ & unitless \\ 
	$\tilde{r}_T, \tilde{r}_P, \tilde{r}_C$ & min$^{-1}$ & & $\displaystyle {r}_T = \frac{\tilde{r}_T}{\tilde{r}_S}, \,\,\, {r}_P = \frac{\tilde{r}_P}{\tilde{r}_S}, \,\,\, {r}_C = \frac{\tilde{r}_C}{\tilde{r}_S}$ & unitless \\ 
	$\tilde{K}$ & $\displaystyle\frac{\text{CFU}}{\text{mL}}$ &  & $\displaystyle {K} = \frac{\tilde{K}}{\tilde{K}} = 1$ & unitless \\  
	$\tilde{\eta}_T, \tilde{\eta}_C$ & $\displaystyle\frac{\text{mL}}{\text{PFU min}}$  &  & $\displaystyle {\eta}_T = \frac{\tilde{\eta}_T \tilde{K}}{\tilde{r}_S}, \,\,\, {\eta}_C = \frac{\tilde{\eta}_C \tilde{K}}{\tilde{r}_S}$ & $\displaystyle\frac{\text{CFU}}{\text{PFU}}$ \\ 
	$\tilde{\gamma}_T, \tilde{\gamma}_I, \tilde{\gamma}_P, \tilde{\gamma}_C$ & min$^{-1}$ &   & $\displaystyle {\gamma}_T = \frac{\tilde{\gamma}_T}{\tilde{r}_S}, \,\,\,  {\gamma}_I = \frac{\tilde{\gamma}_I}{\tilde{r}_S}, \,\,\, {\gamma}_P = \frac{\tilde{\gamma}_P}{\tilde{r}_S}, \,\,\, {\gamma}_C = \frac{\tilde{\gamma}_C}{\tilde{r}_S}$  & unitless \\
	$\tilde{\delta}$ & min$^{-1}$ &  & $\displaystyle {\delta} = \frac{\tilde{\delta}}{\tilde{r}_S}$ & unitless \\ 
	$\tilde{f}_T$ & unitless &   & ${f}_T = \tilde{f}_T$  & unitless \\ 
	$\tilde{f}_C$ & unitless &  &  ${f}_C = \tilde{f}_C$  & unitless \\ 
	$\tilde{A}$ & $\mu$g/mL &  &  $A = \tilde{A}$  & $\mu$g/mL  \\
	$\tilde{\kappa}$ & $\displaystyle \frac{\text{mL}}{\mu\text{g min}}$ &  &  $\displaystyle\kappa = \frac{\tilde{\kappa}}{\tilde{r}_S}$  & mL/$\mu$g  \\
	$\tilde{\epsilon}$ & $\displaystyle \frac{\text{mL}}{\mu\text{g min}}$ &  &  $\displaystyle\epsilon = \frac{\tilde{\epsilon}}{\tilde{r}_S}$  & mL/$\mu$g  \\
	$\tilde{k}$ & min$^{-1}$ &  &  $k = \frac{\tilde{k}}{\tilde{r}_S}$  & unitless \\
	$\tilde{\beta}_T$ & $\displaystyle\frac{\text{PFU}}{\text{CFU}}$ &   & ${\beta}_T = \tilde{\beta}_T$ & $\displaystyle\frac{\text{PFU}}{\text{CFU}}$ \\ 
	$\tilde{\beta}_C$ & $\displaystyle\frac{\text{PFU}}{\text{CFU min}}$ &  & $\displaystyle {\beta}_C = \frac{\tilde{\beta}_C}{\tilde{r}_S}$ & $\displaystyle\frac{\text{PFU}}{\text{CFU}}$   \\ 
	$\tilde{\mu}_T, \tilde{\mu}_C$ & min$^{-1}$ &   & $\displaystyle {\mu}_T = \frac{\tilde{\mu}_T}{\tilde{r}_S}, \,\,\, {\mu}_C = \frac{\tilde{\mu}_C}{\tilde{r}_S}$ & unitless \\ 
	\hline
	\end{tabular}
	 \end{center}
	\label{tab:paramfull}
	\end{table}

\subsection{Parameter selection} \label{par}
The \textbf{growth rate} $\tilde{r}_S$ for \textit{P.~aeruginosa in vitro} is approximately 5.1e-3 min$^{-1}$ \cite{spalding2018mathematical}, although \textit{P.~aeruginosa} growth is highly variable in humans \cite{kopf2016trace}. Therefore all rate parameters provided in min$^{-1}$ are divided by this rate in order to rescale (see Table \ref{tab:paramfull}). 

The growth rates for latently infected bacteria, $\tilde{r}_T$ and $\tilde{r}_C$, are assumed to be similar to uninfected bacteria. Our estimate for the growth rate $\tilde{r}_P$ of productive bacteria is based on \textit{E.~coli} infected with the filamentous phage M13 \cite{shapiro2016evolution}; depending on the experimentally imposed fitness pressure, chronic viruses appear capable of either increasing or decreasing host growth rates. In the interest of simplicity, we assumed that chronic bacteria grow at a rate similar to uninfected bacteria.

The \textbf{carrying capacity} $\tilde{K}$ of bacteria in a medium depends on the environment. Even within the sputum of patients with cystic fibrosis, the carrying capacity is difficult to estimate due to variability among patients. One study of patients with cystic fibrosis found that the densities of viable \textit{P.~aeruginosa} in sputum of 12 patients not undergoing treatment ranged from 5.3e3 CFU/mL to 1.8e11 CFU/mL \cite{stressmann2011does}. Because of the wide range of carrying capacities, we use the geometric mean of this range, 3e7 CFU/mL, for illustrative purposes only. Because $\tilde{K}$ is challenging to estimate, we use strategies that do not explicitly require $\tilde{K}$ to estimate the infection rate.

The \textbf{infection rate} for \textit{E.~coli} and T4 phage in mucus (assuming mass action infection) is known to be approximately 47e-10 mL/min per PFU \cite{stent1963molecular,barr2015subdiffusive}. Infection in marine ecosystems (also assuming mass action infection) is similar at around 24e-10 mL/min per PFU \cite{stent1963molecular,thingstad2014theoretical}. Given our uncertainty in the bacterial carrying capacity, we elect to use an infection rate within the range given by Sinha et al. \cite{sinha2017silico}; the authors fit their mass action infection model to time series population data that reached carrying capacity. The authors present $\tilde{K} \tilde{\eta} \in [0.45, 100]$ hr$^{-1}$ and $\tilde{r}_S\in[0.5, 10]$ hr$^{-1}$. Rescaling leads to a range of ${\eta}$ between $0.045$ and $200$, and our selected value of $\eta=1$ is near the geometric mean of that range.

The \textbf{phage production delay rate} $\tilde{\delta}$ is estimated based on the eclipse and rise phase of PAXYB1 and $\varphi$PSZ1 phage \cite{yu2017characterization,el2015isolation}. The eclipse (latent) and rise phase is 130 minutes total for PAXYB1 \cite{yu2017characterization} and 27 minutes total for $\varphi$PSZ1 \cite{el2015isolation}. The smallest (rescaled) delay rate is then $1/130/$5.1e-3$=1.5$, and the largest is $1/27/$5.1e-3$=7.3$. We selected the approximate average of this range, 4, to be the delay rate $\delta$.

The \textbf{temperate lysogen frequency} $f_T$ is estimated based on \textit{E.~coli} and $\lambda$ phage \cite{oppenheim2007new}: ``It is known that a cell infected by one phage predominantly follows the lytic default pathway (about 99\% of the time).''

The \textbf{phage degradation rates} $\tilde{\mu}_T$ and $\tilde{\mu}_C$ were estimated based on the decay rates of phage in aquatic environments \cite{heldal1991production}. The decay rates ranged from 0.26 to 1.1 per hour. We rescale by multiplying by 60 minutes per hour and the bacterial growth rate, 5.1e-3 per minute. The rescaled range of decay rates is then 0.9 to 3.6. We selected a value of $\mu_T = \mu_C = 1$ arbitrarily from this range.

The \textbf{amplitude of stress} $A$ is based on antibiotic concentrations ($\mu$g/mL) in blood serum \cite{grillon2016, fong1986}. The antibiotics of interest in these studies were ciprofloxacin and levofloxacin. 

The \textbf{metabolic decay rate of antibiotics} $k$ is calculated based on the half-life of levofloxacin \cite{zhanel2006review}. The half-life of a standard dose of levofloxacin within a human is approximately $T_{1/2}$ = 6.9 hours = 414 minutes; the metabolic decay rate is then $\tilde{k} = \ln{(2)}/T_{1/2} \approx 1.7$e-3 min$^{-1}$; rescaling by the growth rate $\tilde{r}_S=5.1$e-3, we get $k\approx 0.3$.

The values of $h_{\beta}$ and $h_{\gamma}$ were selected to be 1 for simplicity. This means the antibiotic concentration for which the production rate is halfway between the minimum and maximum is 1 $\mu$g/mL. Similarly, the antibiotic concentration for which the growth rate is half the maximum is 1 $\mu$g/mL.

All other parameter baselines and ranges are educated guesses.

\subsection{Sensitivity analysis: technical details}
In all analyses, we have used the initial conditions $S (0) = 1\mathrm{e}{-3}, V_T (0) = V_C (0) = 1\mathrm{e}{-7}$, following Sinha et al. \cite{sinha2017silico}.

The global uncertainty and sensitivity analysis was performed using the methodology outlined in Marino et al. \cite{marino2008methodology}. The base code for the analysis is freely available at the author Denise Kirschner's website \cite{marinoCode}. 

In brief, the analysis uses Latin Hypercube Sampling (LHS) of parameter space to simulate uncertainty in model parameters. LHS sampling requires fewer model simulations than simple random sampling without introducing bias \cite{mckay1979comparison}. We used uniform sampling of each parameter about the base values given in Table II. The ranges of the uniform samples are available in our code.

We use Partial Rank Correlation Coefficients (PRCC) to test the sensitivity of model outputs to parameter uncertainty because model outputs generally depend monotonically on model inputs, but those relationships are not linear trends. As noted by Marino et al. \cite{marino2008methodology}, for linear trends we could have used Pearson correlation coefficient (CC), partial correlation coefficients (PCCs), or standardized regression coefficients (SRC). Had our trends been non-monotonic, we would have used the Sobol method or one of its many extensions \cite{saltelli2002making}.
 
The displayed sensitivities in Fig 7 are the PRCCs for the model output $y$ (either minimum $T$ or minimum $\kappa$) and the model inputs $x_j$. As described by Marino et al. \cite{marino2008methodology}, partial rank correlation characterizes the monotonic relationship between input $x_j$ and output $y$ after the effects on $y$ of the other inputs are removed. The values of PRCCs fall between $-1$ and $1$, with $1$ indicating the strongest positive rank correlation and $-1$ indicating the strongest negative rank correlation. The significance indicates the probability that the rank correlation is zero (i.e., large significance suggests that there is no relationship between $x_j$ and $y$).

\end{document}